\documentclass[12pt,a4paper]{article}
\usepackage{epsfig}

\begin{document}

\begin{flushright}
UA/NPPS-5-04\\
\end{flushright}

\begin{center}
{\LARGE{\bf{Prospects of detecting the QCD critical point}}}\\
\vspace{0.5cm}
N.~G. ANTONIOU, Y.~F. CONTOYIANNIS, F.~K. DIAKONOS\\ 
and A.~S. KAPOYANNIS\\

{\it{Department of Physics, University of Athens,\\
15771 Athens, Greece}}\\
\end{center}

\begin{abstract}
We investigate the possibility to observe the QCD critical point
in $A+A$ collisions at the SPS. Guided by the QCD phase diagram 
expressed in experimentally accessible variables we suggest that the 
process $C+C$ at $158~GeV/n$ freezes out very close to the critical  
point. We perform an analysis of the available preliminary 
experimental data for a variety of SPS processes. The basic tool
in our efforts is the reconstruction of the critical
isoscalar sector which is formed at the critical point.
Our results strongly support our proposition regarding the $C+C$ 
system.
\end{abstract}


\section{Critical properties of QCD}

The study of the QCD phase diagram in the baryonic chemical 
potential-temperature plane is a subject of rapidly increasing
interest in the last decade. Recent investigations \cite{SRS} 
suggest that in the real world where the $u$ and $d$ quarks have a 
small current mass ($O(10 MeV)$) and the strange quark is much 
heavier ($O(100 MeV)$) there is a second order critical point as 
endpoint of a first order transition line. This critical endpoint is 
located at low baryonic density (compared to the baryonic density in 
the nuclear matter) and high temperature ($O(100 MeV)$) values. The
order parameter characterizing the critical behaviour has isoscalar
quantum numbers and the underlying symmetry which breaks 
spontaneously at the critical point is the $Z(2)$ symmetry 
classifying the QCD critical point in the $3-D$ Ising
universality class \cite{Ising}. However this symmetry does not
represent an obvious symmetry of the original QCD Langrangian 
\cite{KLS} but it is rather an invariance of the effective thermal QCD
action.

The fluctuations of the condensate formed at the 
critical point correspond to isoscalar particles which are 
distributed in phase space producing a characteristic self-similar 
pattern with fractal geometry determined by the isothermal critical
exponent of the $3-D$ Ising universality class \cite{Ant1}. The 
properties of the isoscalar condensate $\sigma(\vec{x})$ are 
strongly affected by the baryonic environment:
\begin{equation}
\sigma_{\rho} \approx \lambda \left( \frac{\rho - \rho_c}{\rho_c} \right)
\sigma_o
\label{eq:eq1}
\end{equation}
where $\rho$ is the baryonic density in the critical region, $\rho_c$ is the 
critical baryonic density and $\lambda$ is a dimensionless parameter of
order one. Eq.(\ref{eq:eq1}) relates the isoscalar condensate at zero
baryonic density ($\sigma_o$) with its value at baryonic density $\rho$.
The form of eq.(\ref{eq:eq1}) suggests that the difference $\rho-\rho_c$ can be 
considered as an alternative order parameter (besides the isoscalar
condensate $\sigma$) characterizing the QCD critical point. Projecting the 
baryonic density onto the rapidity space and using the scaling properties of
the critical baryonic fluid formed in a $A+A$-collision process,
one obtains the relation \cite{Ant2}:
\begin{equation}
A_{\perp}^{-2/3} n_b = \Psi(z_c,\frac{\mu}{\mu_c},\rho_c)
\label{eq:eq2}
\end{equation}
where $A_{\perp}$ is the total number of nucleons of the $A+A$ system in the plane 
transverse to the beam, $n_b$ is the net baryon density at midrapidity and $\Psi$ 
is a scaling function. The variable $z_c$ is defined as: $z_c = A_{\perp}^{-2/3} 
A_t L^{-1}$ with $A_t$ the total number of participating nucleons
in the $A+A$ collision and $L$ the size of the system in rapidity space.
The scaling function $\Psi$ depends also on the ratio of the chemical potentials
$\frac{\mu}{\mu_c}$ and for $\mu=\mu_c$ simplifies to:
$$\Psi(z_c,1,\rho_c) =\left\{ \begin{array}{c} \rho_c +
\frac{2}{\pi} (z_c - \rho_c) + C (z_c -\rho_c)^4~~~;~~~z_c > \rho_c \\
z_c~~~;~~~z_c \leq \rho_c \end{array} \right.$$
In fact the scaling relation (\ref{eq:eq2}) represents an alternative description
of the QCD phase diagram in terms of measurable quantities \cite{Ant2}. 
In Fig.~1
we present a plot of eq.(\ref{eq:eq2}) in the $(z_c,\xi)$ plane (we use the notation
$\xi=A_{\perp}^{-2/3} n_b$). In the same plot we show also the coordinate pairs 
$(z_{c,i},\xi_i)$ for a variety of $A+A$ processes in running (NA49, RHIC), 
passed (NA35) and future experiments (LHC). We also mark the QCD critical point in
this graph. One can easily see that the $C+C$ system at the SPS energies 
($158~GeV/n$) is very close to the critical point. 

\begin{figure}[ht]
\centerline{\epsfxsize=5.9in\epsfbox{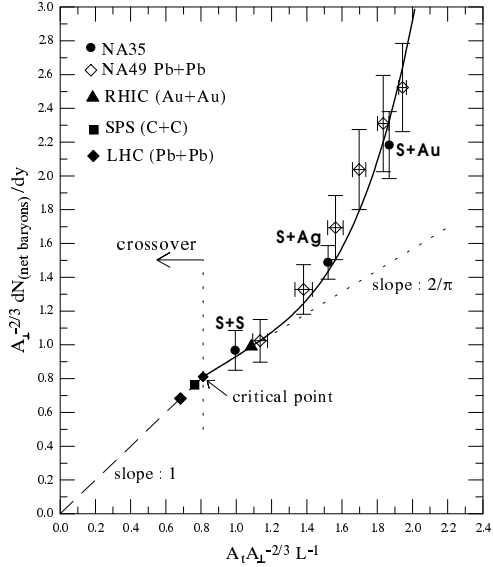}}
\vspace{-12cm}   
\caption{The QCD phase diagram in experimentally accessible variables
according eq.~(2). The various processes in recent and future heavy-ion
collision experiments are also displayed.
\label{fig1}}
\end{figure}

Notice that the remaining 
$A+A$ processes at the SPS ($Si+Si$, $Pb+Pb$) are not so close to the critical
point although they still lie in the scaling region. It is therefore expected that 
nonstatistical fluctuations will be present in all these processes and become stronger
as we approach the critical point. How to reveal these fluctuations, it will be discussed
in the following sections.  

\section{Statistical description of the isoscalar condensate}

The isoscalar condensate formed at the critical point can be described as
a critical (Feynman-Wilson) fluid in local thermal equilibrium \cite{Ant3}.
Universality class arguments determine the effective action for the dynamics 
of the condensate $\sigma(\vec{x})$ at energies $\approx T_c$ in $3-D$ as:
\begin{equation}
\Gamma_c[\sigma]=T_c^{-1} \int d^3
\vec{x}\left[\frac{1}{2}(\nabla \sigma)^2
+ GT_c^4 (T_c^{-1} \sigma)^{\delta +1}\right]
\label{eq:eq3}
\end{equation}
Eq.(\ref{eq:eq3}) leads to the correct equation of state:
$${\displaystyle{\frac{\delta \Gamma}{\delta \sigma}}}
\sim G \sigma^{\delta}$$
where $\delta =5$ is the isothermal critical exponent of the $3-D$
Ising model.
The coupling $G$ has been calculated in \cite{MT} for the $3-D$ Ising
model on the lattice leading to $G \approx 2$. The field $\sigma(\vec{x})$
in eq.(\ref{eq:eq3}) is in fact macroscopic, i.e. the quantum fluctuations
are integrated out to get the effective action (\ref{eq:eq3}) and therefore
it possesses classical properties. Following ref.(5) we recall here that 
the experimentally accessible quantity is not the field 
$\langle \sigma \rangle$ itself but 
the quantity $\langle \sigma^2 \rangle$ which represents  
density fluctuations of the $\sigma$-particles created at $T=T_c$.

Based on the effective action (\ref{eq:eq3}) we can now proceed to determine
the partition function of the condensates as a functional integral:
\begin{equation}
Z=\int {\mathcal{D}} [\sigma] e^{-\Gamma_c[\sigma]}
\label{eq:eq4}
\end{equation}
The path summation in the above equation is dominated by the saddle points 
of the corresponding action which have an instanton-like form \cite{Ant1}.
Within this approximation we can determine the density-density correlation
of the critical system both in configuration as well as in momentum space.
Then using the calculated density-density correlation function we find the 
distribution of the corresponding $\sigma$-particles in phase space. We end up 
with a pattern formed through the overlapp of several self-similar clusters with 
fractal mass dimension determined by the isothermal critical exponent $\delta$
\cite{Ant1}. 
Using the Fourier transform of the spatial density-density correlation function
we obtain the corresponding quantity in momentum space. A similar pattern occurs
also in momentum space. The number of clusters as well as the multiplicity
within each cluster are the same in both spaces while the local fractal dimension
differs. Another property determining the geometrical features of the critical 
system is the shape of its evolution. For a cylindrical evolution the number of
clusters is in general greater than one while in the case of spherical evolution
the system consists a single cluster. Also the corresponding fractal dimensions 
are influenced by the geometrical shape of the evoluting system \cite{Ant3}. 
A less influenced property is the fractal dimension of the transverse momentum
space which turns out to be $\approx 0.7$ for cylindrical systems and 
$\approx 1$ for spherical systems.
\begin{center}
\noindent{\bf{The Critical Monte Carlo (CMC) event generator}}
\end{center}
Using the results of the saddle point approximation to the partition function
of the critical system one can develop a Monte-Carlo algorithm to simulate the
production of the critical $\sigma$-particles in an $A+A$ collision. 
We restrict our interest to the
the distribution of the sigmas in momentum space since the coordinates in this
space are experimentally accessible. The momentum coordinates of the centers of 
the $\sigma$-clusters are treated as random variables distributed according to 
an exponential decay law with range determined by the critical temperature. 
Within each cluster 
the particles are strongly correlated and possess a fractal geometry. The 
corresponding fractal dimension is given in terms of the exponent $\delta$ while
the multiplicity within each cluster is determined through the transverse radius 
of the entire system, its size in rapidity, the critical coupling $G$ and the 
critical temperature $T_c$. The momenta of the sigma-particles within each cluster
are generated using the tool of L\'{e}vy walks \cite{AZ}.

Exactly at the critical temperature $T_c$ the mass of the sigma particles is zero
(for an infinite system). As the system freezes out the sigma-mass increases and when it overcomes the
two pion threshold the sigmas decay into pions which constitute the experimentally
observable sector of the critical system. Unfortunately there is no theoretical
description of this process based on first principles. A possible treatment of the
$\sigma$-decay into pions is to introduce the probability density $P(m)$ for a sigma
to have mass $m$ and then using pure kinematics to determine the momenta of the 
produced pions. The mass $m$ is assigned to the decaying sigmas randomly. A more
detailed description of the whole algorithm can be found in \cite{Ant3}.
  
\section{SPS data analysis (preliminary)}

If the mass of the decaying sigma is well above the two-pion threshold the 
momenta of the produced pions are very distorted with respect to the momentum 
of the initial sigma and the fractal geometry of the critical condensate is 
not transfered to the pionic sector. Therefore in order to reveal the critical
fluctuations in an analysis of the final pions one has to isolate the part of 
phase space for which the mass of the decaying sigmas is very close to the 
two-pion threshold. In this case the fractal properties of the sigma-momenta 
are transfered to final pions. Our proposal is to perform an event by event 
analysis in an $A+A$-dataset forming for each event all the pairs of pions 
with opposite charge and filtering out those pairs with invariant mass within 
a narrow window just above the value $2 m_{\pi}$ \cite{Ant3}:
\begin{equation}
2 m_{\pi} \leq \sqrt{m^2_{\pi^+ \pi^-}} \leq 2 m_{\pi} + \epsilon~~~;~~~~~
m^2_{\pi^+ \pi^-}=(p_{\pi^+} + p_{\pi^-})^2
\label{eq:eq5}
\end{equation}
In (\ref{eq:eq5}) $p_{\pi^{\pm}}$ are the four-momenta of the positive 
(negative) charged pions respectively. The parameter $\epsilon$ is assumed 
to be very small compared to the pion mass. We apply first our analysis to 
a large set of CMC generated events (100000). The CMC input parameters have 
been chosen to meet the properties of the $C+C$ system at the SPS:

\begin{itemize}
\item{The size in rapidity $\Delta=6$, corresponding to $\sqrt{s}=158~GeV/n$}
\item{The transverse radius $R_{\perp}=15~fm$. With this choice we can fix
the mean multiplicity of charged pions to be $\approx 50$ close to the 
corresponding value in the $C+C$-system at the SPS}
\item{The critical temperature $T_c \approx 140-170~MeV$}
\item{The self-couplig $G=2$ and the isothermal critical exponent $\delta=5$
detrmined by the universality class of the transition}
\end{itemize}
  
The essential parameters in our approach is the transverse radius $R_{\perp}$
which controls the mean pion multiplicity in the simulation of an 
$A+A$-process, the size in rapidity $\Delta$ which controls the total energy 
of the system and the isothermal exponent $\delta$ determining the fractal 
geometry of the isoscalar fluctuations. 

Having produced the $10^{5}$ CMC events we calculate the factorial moments in 
transverse momentum space of the final produced chraged pions. For the decay 
of the $\sigma$-s into pions we use a Gaussian $\sigma$-mass probability 
distribution with a large mean value ($300~MeV$) and large standard deviation 
($100~MeV$). With such a deviation in mass we expect that the critical 
fluctuations present in the sigma-sector will be strongly suppressed in the 
charged pion-sector. This choice of $P(m_{\sigma})$ may be quite conservative 
but it consists a good test for the efficience of our data-analysis algorithm. 

Then using the charged pion momenta of each event one can form $\pi^+ - \pi^-$ 
pairs with invariant mass very close to the two-pion threshold. In our actual 
calculations we have used as a window $\epsilon=4~MeV$ to filter out pion 
pairs with invariant mass in the range 
$2 m_{\pi} \leq \sqrt{m_{\pi^+ \pi^-}^2} \leq 2 m_{\pi} + \epsilon$. 
Then we consider each charged pion pair as a sigma-particle.
Performing the factorial moment analysis in the new momenta (of the 
reconstructed sigmas) we expect a partial restoration of the critical 
fluctuations. Indeed this characteristic behaviour is clearly shown in 
Fig.~2a where we present the results of the calculation of the second 
factorial moment in transverse momentum space for both the negative pions 
as well as the sigmas. The effect of the restoration of the critical 
fluctuations in the reconstructed sigma sector combined with the
large suppression of the fluctuations in the negative pions, as predicted, is 
impressive. The theoretical expectation, for an infinite critical system, for
the corresponding intermittency index is $s_{2,cr}^{(2D)}\approx 0.67$ while our
analysis leads to $s_2^{(2D)}\approx 0.54$. We have applied the same analysis
to data sets obtained from the NA49 experiment at the SPS. We have analysed
13731 events of the $C+C$ system at $158~GeV/n$, 76065 events of the 
$Si+Si$ system at the same energy, 13420 events of the $Pb+Pb$ system at 
$40~GeV/n$, 384 events of the same system at $80~GeV/n$ and finally 5584
events of the $Pb+Pb$ system at $158~GeV/n$. It must be noted that all the
data sets used in our analysis are only preliminary and there is a need for
further investigations with improved data. In Fig.~2b we show the results for 
the second moment in transverse momentum space in the $C+C$ system. There is 
an impressive agreement between simulated and real data. In fact the slope 
$s_2^{(2D)}$ of the $C+C$ system turns out to be $\approx 0.58$.

\begin{figure}[ht]
\centerline{\epsfxsize=4.9in\epsfbox{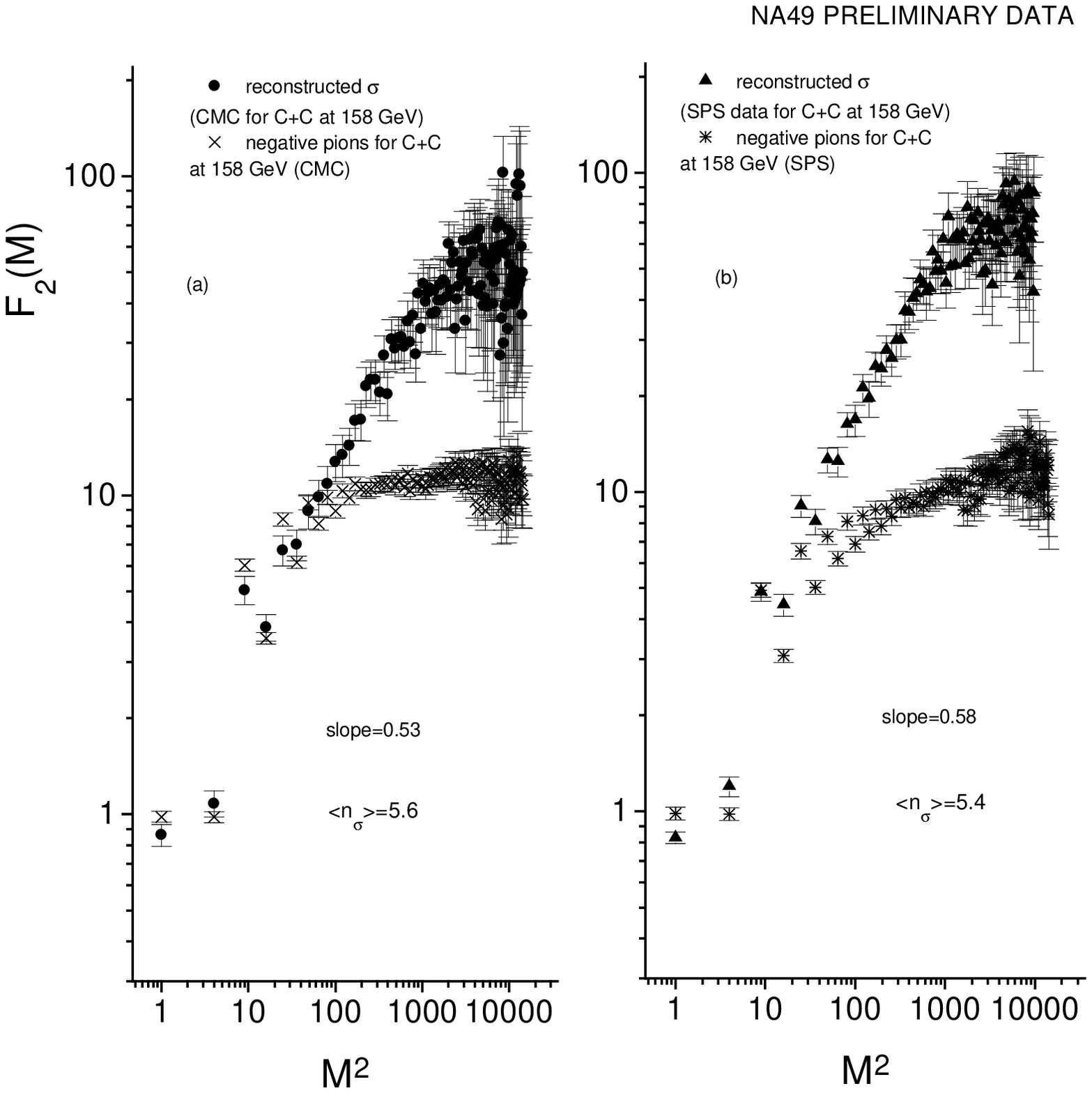}}   
\vspace{-6cm}
\caption{The reconstruction of $\sigma$-momenta in (a) $10^{5}$ CMC 
events and (b) 13731 preliminary $C+C$ data.
\label{fig2}}
\end{figure}

In Fig.~3 we show the second factorial moment for all the available NA49 
experimental data sets both for negative pions as well as sigmas.
A gradual increament of the slope $s_{2}^{(2D)}$ as we approach the 
$C+C$ system - and according to Fig.~1 the critical point - is observed close
to our theoretical expectations. For all the systems the effect of the 
reconstruction of the critical fluctuations in the $\sigma$-sector is clearly
seen. 

\begin{figure}[ht]
\centerline{\epsfxsize=3.9in\epsfbox{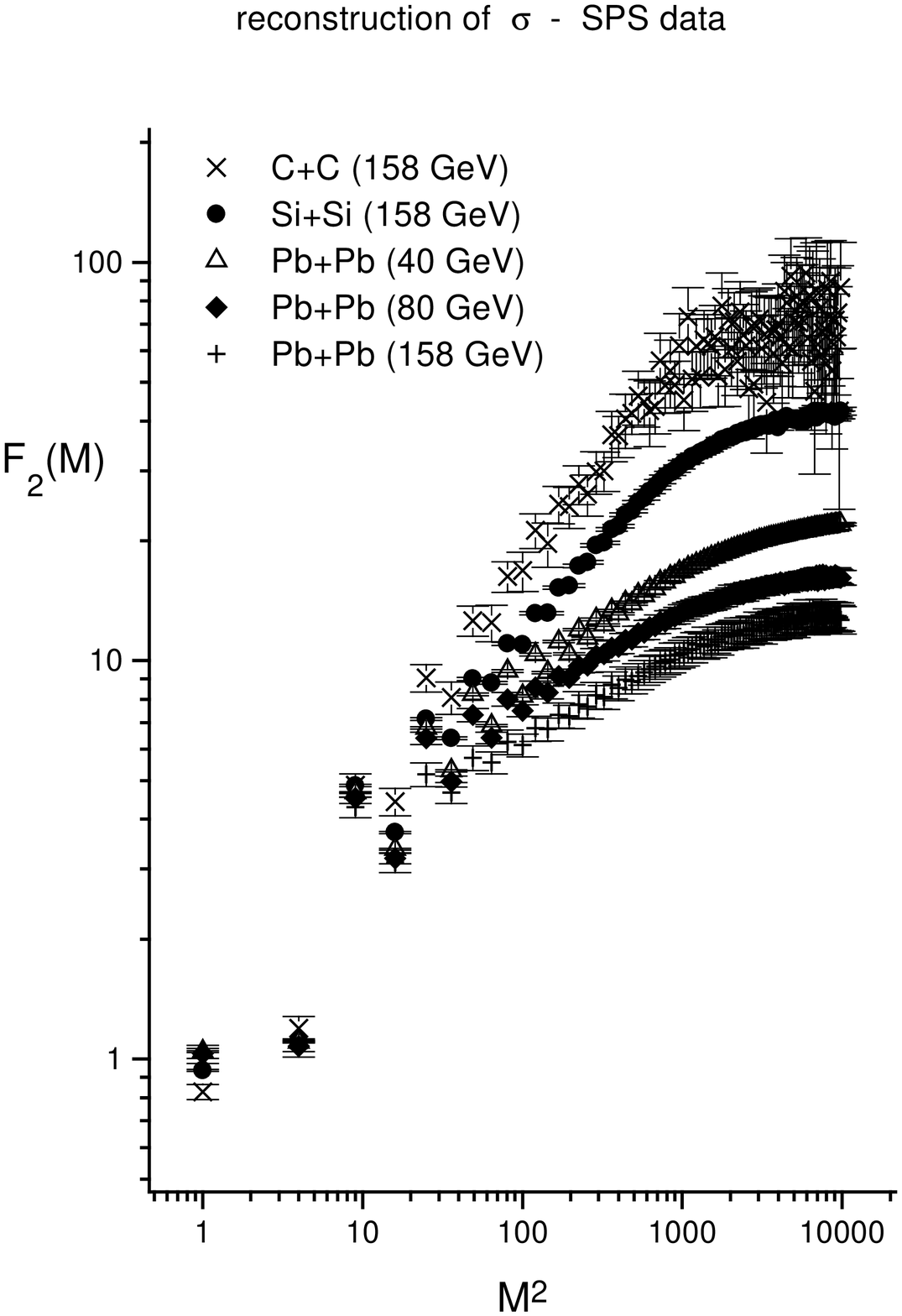}}   
\caption{The second factorial moment in transverse momentum space for 
all the analysed SPS processes. Represented are only the results 
obtained after the reconstruction of the isoscalar sector.
\label{fig3}}
\end{figure}

The analysis described so far concerns a finite kinematic window above
the two pion threshold. It is interesting to extrapolate the properties
of the various systems exactly at the two pion threshold. In this case
no distortion due to the $\sigma$-decay into pions will be present and 
we expect to reproduce the theoretically expected results for the 
critical system. Therefore we have to take the limit $\epsilon \to 0$.
In order to extract this information one has to calculate $s_2^{(2D)}$ for 
various values of the kinematical window $\epsilon$ and use an interpolating
function to extrapolate to $\epsilon=0$. The obtained value 
$s_{2,o}^{(2D)}$ can be directly compared with the theoretical expected
value for $s_{2,cr}^{(2D)}$. To be able to perform this analysis one has to study
a system with very large charged pion multiplicity per event and/or to use
a very large dataset. For this reason we have applied our approach to two 
systems: (i) the 5584 $Pb+Pb$ events at $158~GeV/n$ and (ii) the 
$10^{5}$ CMC generated critical events (simulating the $C+C$ system 
at $158~GeV/n$). The
results of our calculations are presented in Fig~4. The solid circles
are the values of $s_2^{(2D)}$ for the $Pb+Pb$ system while the open 
triangles describe the CMC results for various values of $\epsilon$. 
The dashed lines present a corresponding exponential fit. 

\begin{figure}[ht]
\centerline{\epsfxsize=3.9in\epsfbox{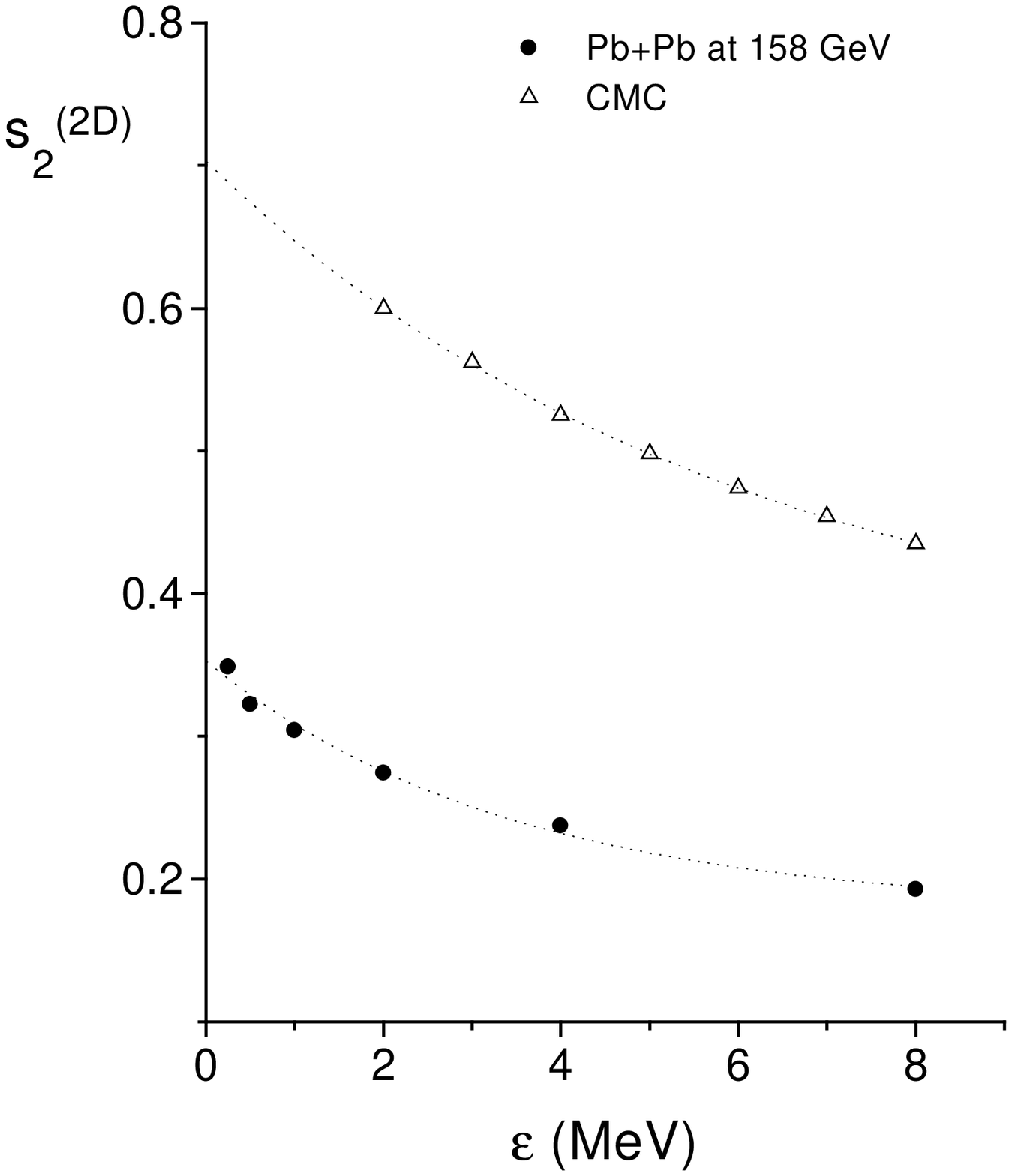}}   
\vspace{-2cm}
\caption{The slope $s_2^{(2D)}$ for different values of the kinematic
window $\epsilon$ both for the $10^{5}$ CMC events as well as for 5584
$Pb+Pb$ events at $158~GeV/n$ using preliminary SPS-NA49 data. 
\label{fig4}}
\end{figure}

For the CMC events
we find $s_{2,o}^{(2D)}=0.69 \pm 0.03$ a value which is 
very close to
the expected $s_{2,cr}=0.67$, while for
the $Pb+Pb$ at $158~GeV/n$ system we get $s_{2,o}^{(2D)}=0.34$. 
The last value corresponds to a strong effect, owing to the fact
that the $Pb+Pb$ system lies
in the scaling region around the critical point. However it is clearly 
smaller
than the theoretical value at the endpoint, in accordance with the fact that this
system freezes out in a distance from the critical point in terms of
the variables in Fig.~1.     

In summary we have introduced an algorithm to detect critical fluctuations
related to the formation of an isoscalar condensate in $A+A$-collisions. 
First analysis, using preliminary SPS-NA49 data, indicates 
the proximity to the critical point of the freeze-out area in the 
collisions with nuclei of medium size ($C+C$ or $Si+Si$).

\section*{Acknowledgments}
The authors thank the NA49 Collaboration for supplying the preliminary
experimental data from the SPS.

\end{document}